\documentclass[12pt]{article}%

\begin{document}

\begin{center}{\large\bf de Sitter Symmetry and Quantum Theory}\end{center}
\vskip 1em \begin{center} {\large Felix M. Lev} \end{center}
\vskip 1em \begin{center} {\it Artwork Conversion Software Inc.,
1201 Morningside Drive, Manhattan Beach, CA 90266, USA
(Email:  felixlev314@gmail.com)} \end{center}
\vskip 1em

\begin{abstract}
de Sitter symmetry on quantum level implies that operators describing a given system satisfy 
commutation relations of the de Sitter algebra. This approach gives a new perspective on fundamental
notions of quantum theory. We discuss applications of the approach to the cosmological constant
problem, gravity, and particle theory.
\end{abstract}
\begin{flushright} PACS: 11.30Cp, 11.30.Ly\end{flushright}
\begin{flushleft} Keywords: quantum theory, de Sitter invariance, Galois fields, gravity\end{flushleft}
\section{Introduction: symmetry on quantum level} 
The most well-known way of implementing Poincare invariance on quantum level is quantum field theory (QFT) on
Minkowski space. Here one starts from classical fields on that space and constructs a Lagrangian.
This makes it possible to calculate the four-momentum $P^{\mu}$ and Lorentz angular momenta $M^{\mu\nu}$
($\mu,\nu=0,1,2,3,\quad M^{\mu\nu}=-M^{\nu\mu}$) for the system of fields under consideration. After quantization,
$P^{\mu}$ and $M^{\mu\nu}$ become operators which should satisfy the commutation relations
\begin{eqnarray}
&&[P^{\mu},P^{\nu}]=0\quad [P^{\mu},M^{\nu\rho}]=-i(\eta^{\mu\rho}P^{\nu}-\eta^{\mu\nu}P^{\rho})\nonumber\\
&&[M^{\mu\nu},M^{\rho\sigma}]=-i (\eta^{\mu\rho}M^{\nu\sigma}+\eta^{\nu\sigma}M^{\mu\rho}-
\eta^{\mu\sigma}M^{\nu\rho}-\eta^{\nu\rho}M^{\mu\sigma})
\label{PCR}
\end{eqnarray}
where $\eta^{\mu\nu}$ is the diagonal metric tensor such that
$\eta^{00}=-\eta^{11}=-\eta^{22}=-\eta^{33}=1$. 

The requirement that the relations (1) should be satisfied
is a must in any relativistic quantum theory since {\it they represent the definition of
Poincare symmetry on quantum level}. These relations do not involve Minkowski space at all and should
be valid regardless of whether the operators $(P^{\mu},M^{\mu\nu})$ have been obtained in QFT or in other
approaches. In typical QFTs the relations (1) can be formally checked by using equal-time commutation relations 
between the field operators (see e.g. textbooks [1,2]).
However, the operators $(P^{\mu},M^{\mu\nu})$ in QFT are constructed from products
of interacting fields at the same points and it is well known that such products are not well defined.
A proof that in interacting QFTs it is possible to construct well defined operators $(P^{\mu},M^{\mu\nu})$ satisfying
Eq. (1) is a very difficult unsolved problem.

The idea of symmetry on quantum level follows. Each system is described by a set of independent operators.
By definition, the rules about how these operators commute with each other define the symmetry algebra. For example,
{\it the definition of de Sitter (dS) invariance on quantum level} is that the representation operators
of the dS algebra describing a quantum system under consideration, should satisfy the commutation relations
\begin{equation}
[M^{ab},M^{cd}]=-i (\eta^{ac}M^{bd}+\eta^{bd}M^{ac}-
\eta^{ad}M^{bc}-\eta^{bc}M^{ad})
\label{CR}
\end{equation}
where $a,b=0,1,2,3,4$, $M^{ab}=-M^{ba}$ and $\eta^{ab}$ is the diagonal metric tensor such that
$\eta^{00}=-\eta^{11}=-\eta^{22}=-\eta^{33}=-\eta^{44}=1$. The validity of these relations is a must
in any de Sitter invariant quantum theory, regardless of whether the operators $M^{ab}$ have been obtained
from QFT on dS space or in other approaches. However, to the best of our knowledge, these relations are
not widely discussed in the literature on quantum dS invariant theories. The same is true in the case
of anti de Sitter (AdS) invariant theories where the commutation relations have the same form (2) but
$\eta^{44}=1$. 

It is usually said that Eqs. (1,2) are written in units $c=\hbar=1$ and, as discussed in Refs.
[3,4], such units have a clear physical meaning. Then in the case of dS and AdS symmetries, all the operators 
$M^{ab}$ are dimensionless while in the case of
Poincare symmetry only the operators of the Lorentz algebra are dimensionless while the momentum operators
have the dimension $1/length$. Equation (1) is a special case of Eq. (2) obtained as follows. If $R$ is a 
parameter with the dimension $length$ and the operators $P^{\mu}$ are {\it defined} as $P^{\mu}=M^{4\mu}/R$ 
then in the formal limit $R\to\infty$ one gets Eq. (1) from Eq. (2). This contraction procedure is well 
known. Hence from the point of view of symmetry on quantum level, dS and AdS symmetries
are more natural and general than Poincare symmetry. It is also clear that on quantum level dS and AdS 
theories can be constructed
without parameters having the dimension of length. Such parameters may be used if one wishes to interpret the
results in classical approximation on dS or AdS space or in the Poincare limit, but they are not fundamental.
In particular, if we accept dS or AdS symmetry on quantum level, then neither the cosmological constant (CC)
$\Lambda=3/R^2$ nor the gravitational constant $G$ can be fundamental (see Refs. [3,4] for a detailed
discussion). 

The problem arises how an elementary particle should be defined. A discussion of numerous
controversial approaches can be found, for example in Ref. \cite{Rovelli}. In the spirit of QFT, fields are more
fundamental than particles and some authors even claim that particles do not exist.
From the point of view of QFT, a possible definition follows \cite{Wein1}: 
{\it It is simply a particle whose field appears in the Lagrangian. 
It does not matter if it is stable, unstable, heavy, light. If its field appears in the Lagrangian
then it is elementary, otherwise it is composite.} We believe that since Eqs. (1) and (2) are treated as
a definition of symmetry on quantum level, the most 
general definition, not depending on the choice of the classical background and on whether we consider 
a local or nonlocal theory, is  that a particle is elementary if the set of its wave
functions is the space of an irreducible representation (IR) of the symmetry algebra in the given theory. 
The relation between the above definitions is discussed in Sec. 4. Note that
the construction of IRs is needed not only for describing elementary particles but even for describing
the motion of a macroscopic body as a whole. For example, when we consider the interaction between two
macroscopic bodies such that the distance between them is much greater than their sizes, it suffices to
describe each body as a whole by using the IR with the corresponding mass.  

\section{de Sitter symmetry and the cosmological constant problem} The data on the cosmological 
acceleration are interpreted in such a way that with the accuracy better than 5\% the value of the CC
is positive. Efforts to explain the value of the CC in the framework of quantum gravity have not been
successful yet, and this problem is as well known as the CC problem. In the literature the existing data
are often explained as a manifestation of dark energy or other fields. The philosophy of such approaches
is roughly as follows: In the absence of matter the spacetime background should be flat, so its
curvature is caused by a hidden matter. However, the notion of the empty spacetime background is not
physical (see e.g. the discussion in Refs. [3,4]). From the point of view of quantum theory, the question is
not whether the empty space is flat or curved but what symmetry algebra is most pertinent for describing
nature. We are not claiming that the dS or AdS algebra is a universal symmetry algebra but at least in
view of the above discussion, each of them is more relevant than the Poincare algebra. 
As noted above, in theories based on the dS or AdS algebra the quantity $\Lambda$ is not fundamental.
As argued in Refs. [3,4], the value of the dimensionful parameter $\Lambda$ simply
reflects the fact that we want to measure distances in meters. Therefore a question why $\Lambda$ is as
it is does not have a fundamental physical meaning. 

Consider a system of two free bodies in dS invariant theory. The motion of each body as a whole is
described by the IR of the dS algebra with the corresponding mass, and the fact that the bodies are
free means that each two-body operator $M^{ab}$ is a sum of the corresponding single-body operators.
Then the result of calculations [3,4] is that in semiclassical approximation
the relative acceleration describing their repulsion is ${\bf a}= \Lambda c^2 {\bf r}/3$ where ${\bf r}$ 
is the vector of the relative distance between the particles. From the formal point of view the result is
the same as in general relativity (GR) on dS space. However, our result has been obtained by using only
standard quantum-mechanical notions while dS space, its metric, connection, etc. have not
been involved at all. We believe this result is a strong indication that the results of GR can be recovered
from semiclassical approximation in quantum theory without using spacetime background and differential
geometry at all. In any case, our result shows that the CC problem does not exist and the phenomenon of 
cosmological acceleration can
be naturally explained without involving dark energy or other unknown fields. The fact that $\Lambda>0$ should be
interpreted not such that the spacetime background is the dS space but that the dS algebra is more
relevant than the Poincare or AdS ones (in which cases one would have $\Lambda=0$ or $\Lambda < 0$, 
respectively).

\section{dS symmetry and gravity} The mainstream approach to gravity is that this phenomenon is a manifestation
of a graviton exchange. The data on binary pulsars are often treated as an indirect indication of the existence
of gravitons but their direct detection has not been successful yet. In recent years a number of works has appeared
where gravity is treated as an emergent phenomenon. We believe that until the nature of gravity has been unambiguously understood, different possibilities should be investigated. dS invariance opens a new approach for investigating
gravity. In our opinion, this approach is clear and natural and the main idea is as follows. 

Consider a spectrum
of the mass operator for a free two-body system in dS invariant theory. This spectrum has been investigated in
Refs. [7,8,3,4]. In contrast to the situation in Poincare and AdS theories where the mass operator is positive
definite and its spectrum is bounded below by $m_1+m_2$ (where $m_1$ and $m_2$ are the masses of the bodies), 
the spectrum of the mass operator in dS theory is not bounded below by this value. Therefore in principle there
is no problem to indicate two-body wave functions for which the mean value of the mass operator contains an
additional term $-Gm_1m_2/r$ with possible corrections given by GR or other classical theories of gravity. 
Here $r=|{\bf r}|$ while $G$
is not a constant taken from the outside but a quantity which should be calculated. The problem is to
understand whether such wave functions are semiclassical and why they are more preferable than other wave
functions. Such a possibility has been first indicated in Ref. \cite{JMP}.

As noted in the preceding section, a standard quantum-mechanical calculation in semiclassical approximation
shows that the relative acceleration of two bodies in dS theory is repulsive and proportional to $r$, i.e. not 
attractive and proportional to $1/r^2$ for gravity as one would expect. In this connection
we note the following. Since all the dS operators are conventional or hyperbolic rotations, the distances in
dS theory should be given in terms of dimensionless angular variables. The angular distance $\varphi$
and the standard distance $r$ are related as $\varphi=r/R$ [4]. It is well known that semiclassical approximation
in quantum mechanics cannot be applied for calculating quantities which are very small. If the distance between two
bodies is large then the angular distance $\varphi$ is not anomalously small and can be calculated in
semiclassical approximation. However, the distances between bodies in the Solar System are much less than $R$
and therefore the angular distances between them are very small if $R$ is very large.

In Ref. [4] it has been argued that standard semiclassical approximation does not apply for macroscopic bodies 
in the Solar system and that the standard distance operator should be modified. We have given a modification,
such that the distance operator has correct properties and semiclassical approximation can be applied. As a
result, the classical nonrelativistic Hamiltonian is
\begin{equation}
H({\bf r}, {\bf q}) =\frac{{\bf q}^2}{2m_{12}} - const\frac{m_1m_2R}{(m_1+m_2)r}
(\frac{1}{\delta_1}+\frac{1}{\delta_2})
\label{preNewton2}
\end{equation}
where ${\bf q}$ is the relative momentum, $m_{12}$ is the reduced mass, $const$ is of order unity
and $\delta_i$ ($i=1,2$) is the width of the dS momentum distribution in the wave function of body $i$.
Therefore the Newton gravitational law can be recovered if $const\cdot R/\delta_i = Gm_i$ where $G$ is
a quantity which should be calculated. This problem will be discussed in Sect. 5. It has also been
shown that the proposed modification naturally gives a correct value for the precession of
Mercury's perihelion. We also discuss whether this approach can reproduce well-known results of GR for the
gravitational red shift of light and the deflection of light by the Sun.

\section{dS symmetry and particle theory} Standard particle theory is based on Poincare symmetry. Since
dS symmetry becomes Poincare one when $R$ is very large and $R$ is much greater than dimensions of
elementary particles, one might think that considering particle theory with dS symmetry is of no interest.
However, we will see below that dS symmetry sheds new light on fundamental notions of particle theory.

We first consider the two definitions of elementary particles given in Sec. 1. In theories
with Poincare and AdS symmetry, there are two kinds of IRs corresponding to particles. IRs with positive
energies are implemented on the upper Lorentz hyperboloid where the temporal component of the four-velocity
is positive: $v_0=\sqrt{1+{\bf v}^2}$ while IRs with negative energies are implemented on the lower Lorentz hyperboloid 
where this component is negative: $v_0=-\sqrt{1+{\bf v}^2}$. IRs with positive energies are associated with
particles and IRs with negative energies - with their antiparticles. Standard particle theory cannot throw away IRs with
negative energies as unphysical. In this theory, positive and negative energy IRs are combined for constructing
a local field satisfying a covariant equation (e.g. the Dirac field satisfying the Dirac equation) and this
field is used for constructing a Lagrangian. Therefore in QFT the two definitions of elementary particles
are usually equivalent. 

One of the ideas of quantization is to circumvent the problem with negative energies. For simplicity we assume
that there are only discrete states which can be enumerated by an integer $i=1,2,...$. In addition,
we define a quantum number $\epsilon$, which shows whether a state with a quantum number $i$ belongs to the
upper or lower hyperboloid. For example, $\epsilon=\pm 1$ for the upper and lower hyperboloids, respectively.
Let $a(i,\epsilon)$ be the operator annihilating the state with quantum numbers $(i,\epsilon)$ and
$a(i,\epsilon)^*$ be the operator creating the state with such quantum numbers. These operators can
satisfy either commutation or anticommutation relations:
\begin{equation}
\{a(i,\epsilon), a(j,\epsilon')^*\}_{\pm}=\delta_{ij}\delta_{\epsilon\epsilon'}
\label{comm}
\end{equation}
where $\delta_{ij}$ is the Kronecker symbol and $\pm$ refers to the anticommutator and commutator, 
respectively. One can define the vacuum vector $\Phi_0$
such that $a(i,\epsilon)\Phi_0=0\,\, \forall i,\epsilon$. Then the energy operator is
\begin{equation}
E=\sum_{i,\epsilon} E(i,\epsilon) a(i,\epsilon)^*a(i,\epsilon)
\label{energy}
\end{equation}
where $E(i,\epsilon)$ is the energy in the state $(i,\epsilon)$. For theories with Poincare and AdS symmetries,
the sign of $E(i,\epsilon)$ is the same as the sign of $\epsilon$. For example, in Poincare invariant theory,
$E(i,\epsilon)=mv_0(i,\epsilon)$ where $m$ is the particle mass which is assumed to be positive and $v_0$ 
is the value of $v_0(i,\epsilon)$ in the state with quantum numbers $(i,\epsilon)$.

At this point we have only rewritten the usual expression for the energy in terms of secondly quantized operators
and hence the problem of negative energies remains. For example, as follows from Eqs. (\ref{comm}) and
(\ref{energy}), any state $a(i,-1)^*\Phi_0$ has a negative energy. Note that the sign of energy is only a matter
of convention. For example, in Poincare invariant theory, a momentum ${\bf p}$ is measured and then the
energy can be defined as $E=\sqrt{m^2+{\bf p}^2}$ but the definition $E=-\sqrt{m^2+{\bf p}^2}$ is 
possible too. It is important, however that the sign of energy should be the same for all particles. For example,
if one defines $E=\sqrt{m^2+{\bf p}^2}$ for the electron and $E=-\sqrt{m^2+{\bf p}^2}$ for the positron
then for the electron-positron system such that the electron has the momentum ${\bf p}$ and the positron has
the momentum $-{\bf p}$, the total energy and momentum would be zero, which contradicts experiment. Hence
we accept the usual convention that the energy of any particle should be positive.

One might try to circumvent the problem of negative energies by saying that the meaning of the operators
$a(i,-1)$ and $a(i,-1)^*$ should be the opposite for the following reason. If $\Phi_1$ is a state with the
energy $E_1$ then $a(i,-1)\Phi_1$ is a state with the energy $E_1-E(i,-1)$ and 
$a(i,-1)^*\Phi_1$ is a state with the energy $E_1+E(i,-1)$. Hence $a(i,-1)$ can be treated as the 
operator of creation of a state with the positive energy $|E(i,-1)|$ and $a(i,-1)^*$ - as the operator
of annihilation of such a state. This idea can be implemented only if the vacuum state is redefined.
For example, the new vacuum can be defined as
\begin{equation}
\Phi_1=\prod_i a(i,-1)^*\Phi_0 
\label{newvac} 
\end{equation}
and then the new treatment of the operators $a(i,-1)$ and $a(i,-1)^*$ is in the spirit of Dirac's hole theory.
However, in that case a new problem arises: as it follows from Eq. (\ref{energy}), the energy of the state
$\Phi_1$ is 
\begin{equation}
E_1=\sum_i E(i,-1)
\label{E1}
\end{equation}
and this is an infinite negative value. It is believed that in quantum gravity the infinite value of the
vacuum energy is unacceptable.

The idea that creation of a state with a negative energy can be described as annihilation of a state with
a positive energy and that annihilation of a state with a negative energy can be described as creation of a state with
a positive energy can also be implemented as follows. Instead of $a(i,-1)$ and $a(i,-1)^*$, define new
operators $b(i)$ and $b(i)^*$ such that $b(i)$ is proportional to $a(i,-1)^*$ and $b(i)^*$ is proportional 
to $a(i,-1)$. For example, if $b(i)=\eta(i)a(i,-1)^*$ where $\eta(i)$ is a complex number then
$b(i)^*=\eta(i)^*a(i,-1)$. These operators will satisfy the same commutation or anticommutation relations
as in (\ref{comm}) if 
\begin{equation}
\eta(i)\overline{\eta(i)}=\pm 1
\label{eta}
\end{equation}
for the case of anticommutators and commutators, respectively [here $\overline{\eta(i)}$ is the complex 
conjugation of $\eta(i)$]. In standard theory (over complex numbers) only the plus sign is possible. 
We now wish to treat $b(i)$ as the 
operator of annihilation of a state with a
positive energy and $b(i)^*$ - as the operator of creation of such a state. Therefore the 
vacuum  state $\Phi$ should now be defined such that $a(i)\Phi=b(i)\Phi=0\,\,\forall i$ where $a(i)\equiv a(i,1)$.
Such a transformation is called the Bogolubov transformation. In that case, if $E(i)\equiv E(i,1)$ and
$E(i,-1)=- E(i,1)$ then, as it follows from Eq. (\ref{energy}), the energy operator can be written as
\begin{equation}
E=\sum_i E(i)\{a(i)^*a(i)\pm b(i)^*b(i)\}\pm E_1
\label{energy2}
\end{equation}
for the case of anticommutation and commutation relations respectively. Here $E_1$ is given by Eq. (\ref{E1}).
We see that if the operators $a$ and $b$ are obtained from the Bogolubov transformation then energies of
antiparticles can be positive only in the case of anticommutation relations. Also, the price for performing
the Bogolubov transformation is the appearance of the infinite constant in Eq. (\ref{energy2}). This
constant is usually neglected by requiring that from the beginning the operators of physical quantities 
should be written in the normal form (when the annihilation operators precede the creation ones). However,
this is an extra requirement which does not follow from the theory. Another way of avoiding the problem
of infinite constants is not to use the Bogolubov transformation at all and require from the beginning
that, regardless of the type of the commutation relations, the energy operator should be  written in the form $\sum_i E(i)[a(i)^*a(i)+b(i)^*b(i)]$.

If a particle is characterized by an additive quantum number (e.g. the electric charge,
the baryon or lepton quantum number) then, since $b^*$ is proportional to $a$, the antiparticle is characterized
by the opposite quantum number. Therefore the sets $(a,a^*)$ and $(b,b^*)$ are independent.
However, in the case of a neutral particle, when all additive quantum numbers are zero, one requires
that the corresponding field be Hermitian. Then the operators $(b,b^*)$ are obsolete
and the number of states describing a neutral field is by a factor of 2 less than the number of states for a
non-neutral field. 

All of the above facts can be found in practically every textbook on QFT. We have mentioned these facts in order
to compare the results of standard theory with those obtained with dS symmetry.
The assumption that quantum theory should be based on dS symmetry implies several far reaching consequences.
First of all, in contrast to Poincare and AdS symmetries, the dS one does not have a supersymmetric
generalization. There is no doubt that supersymmetry is a beautiful idea. On the other hand, one
might say that there is no reason for nature to have both, elementary fermions and elementary bosons since the latter can
be constructed from the former. A well-known historical analogy is that the simplest covariant equation is not the Klein-Gordon
equation for spinless fields but the Dirac and Weyl equations for the spin 1/2 fields since the former is the equation of the
second order while the latter are the equations of the first order.

A very elegant description of IRs of the dS group can be found in a book \cite{Mensky} on the theory of induced
representations for physicists. A crucial difference between dS symmetry on one hand and Poincare or AdS 
symmetry on the other is that in the
dS case one IR can be implemented only on the both, upper and lower Lorentz hyperboloids simultaneously. 
Only in the formal limit $R\to\infty$ one IR of the dS algebra splits into two independent IRs of the Poincare 
algebra on the upper and lower Lorentz hyperboloids [3]. When $R$ is finite, transitions between the 
hyperboloids are not prohibited since the states on the upper and lower hyperboloids belong to the same IR. 

As shown in Ref. \cite{Mensky}, there exists an equivalent description when an IR is implemented not on two 
hyperboloids but on the three-dimensional unit sphere $S^3$ in the four-dimensional space. The points of $S^3$ are characterized by $u=({\bf u}, u_4)$
such that $u_4^2+{\bf u}^2=1$. The relation between the points of the upper hemisphere ($u_4>0$) 
and the upper hyperboloid is ${\bf u}={\bf v}/v_0$ and $u_4=(1-{\bf u}^2)^{1/2}$ while the relation between the points of the lower hemisphere ($u_4<0$) and the lower hyperboloid is ${\bf u}=-{\bf v}/v_0$ and $u_4=-(1-{\bf u}^2)^{1/2}$.
The equator of $S^3$ where $u_4=0$ has measure zero with respect to the upper and lower hemispheres.

In variables $({\bf u}, u_4)$, transitions between the hyperboloids correspond to crossing the equator of $S^3$
and hence such transitions are not singular. The possibility
of such transitions shows that if $R$ is finite then the very notions of a particle and its antiparticle can be 
only approximate and such quantum
numbers as the electric charge, the baryon and lepton quantum numbers cannot be strictly conserved. 
The nonconservation of the baryon and lepton quantum numbers
has been already considered in models of grand unification but
the electric charge has been always believed to be a strictly
conserved quantum number. The
experimental facts that all those numbers are conserved might be a consequence of the fact that nowadays 
the value of $R$ is very
large and probabilities of transitions particle$\leftrightarrow$antiparticle are very small. However, at earlier
stages of the Universe, when $R$ was not so large, those probabilities were not negligible. One might speculate that this
was the reason of the observed baryon asymmetry of the Universe. It is also immediately clear that in the dS case there are 
no neutral particles since it is not possible to reduce the number of states in an IR.

Consider now the problem of quantization in dS theory. We can take $M^{40}$ as the dS analog of the Hamiltonian
since $M^{40}/R$ becomes the Hamiltonian in Poincare limit. By analogy with standard theory, we can define the
operators $a(i,\epsilon)$ satisfying Eq. (\ref{comm}) and the vacuum state $\Phi_0$. Then the energy operator
can be again written in the form (\ref{energy}) [8,3]. In contrast to the situation in standard theory, one
cannot now guarantee that $E(i,1)>0,\,\,E(i,-1)<0\,\,\forall i$. However, this is at least the case for those $i$
when Poincare approximation works with a high accuracy [8,3]. Hence the problem of negative energies exists in
the dS case as well. By analogy with standard theory, one might try to redefine the vacuum as in
Eq. (\ref{newvac}) but this vacuum also will have an infinite energy given by Eq. (\ref{E1}).

One might define the operators $(b,b^*)$ in the same wave as in standard theory
and then, by analogy with standard theory, one gets Eq. (\ref{energy2}) [8,3]. However, since transitions
$particle\leftrightarrow antiparticle$ are not prohibited, in contrast to standard
theory, the Bogolubov transformation in the dS case can be performed only at the expense of breaking dS symmetry [3].
Symmetry breaking will occur only at extremely large energies and in that case the transitions 
$particle\leftrightarrow antiparticle$ will be prohibited after the transformation. If this scenario 
is acceptable and for some
reason (see Sec. 5) the infinite constant $E_1$ given by Eq. (\ref{E1}) can be neglected then we come to
conclusion that in dS theory only fermions can be elementary. We believe, 
however, that since we treat Eq. (2) as a must, we should consider scenarios when difficulties can be resolved 
without breaking symmetry. One of such scenarios is discussed in the next section.

The fact that in Poincare and AdS theories a particle and its antiparticle are described by
different IRs means that they are different objects. Then a problem arises why they have the same masses and spins but
opposite charges. In QFT this follows from the CPT theorem which is a consequence of locality since {\it we construct}
local covariant fields from a particle and its antiparticle with equal masses. A question arises what happens if locality
is only an approximation: in that case the equality of masses, spins {\it etc.}, is exact or approximate? Consider a simple
model when electromagnetic and weak interactions are absent.
Then the fact that the proton and the neutron have the same
masses and spins has nothing to do with locality; it is only a
consequence of the fact that the proton and the neutron belong
to the same isotopic multiplet. In other words, they are simply
different states of the same object---the nucleon. We see, that
in dS invariant theories the situation is analogous. The fact
that a particle and its antiparticle have the same masses and
spins but opposite charges (in the approximation when the
notions of particles, antiparticles and charges are valid) has
nothing to do with locality or nonlocality and is simply a
consequence of the fact that they are different states of the
same object since they belong to the same IR.

Another consequence of dS symmetry is as follows. In QFT a particle and its
antiparticle should be combined into one object, which is a local field.
For example, the Dirac field combines the electron and positron together.
However, in dS theory, Dirac's idea of combining a particle and its antiparticle
together is already implemented since they belong to the same IR. This poses a problem 
whether for constructing quantum theory local fields are needed at all. 

\section{A quantum theory over a Galois field}  

In the preceding sections we discussed symmetries in standard approach to quantum theory, i.e. that quantum
states are represented as vectors in complex Hilbert spaces and operators of physical quantities - as
operators in such spaces. In Ref. \cite{GFQT} we have proposed an approch when  quantum
states are represented as vectors in spaces over a Galois field and operators of physical quantities - as
operators in such spaces. We believe that this approach, which we call a quantum theory over a Galois field
(GFQT), is more elegant and natural than standard approach. A detailed motivation can be found e.g. in
Refs. \cite{11,gravity}. Since any Galois field is finite, in GFQT infinities cannot exist in principle.
One of the motivations of GFQT is that the notion of infinitely small is based on the macroscopic experience
that every macroscopic object can be divided into any number of parts. However, in view of the existence of
atoms and elementary particles, it is clear that standard division has a limited applicability.

Any Galois field $F_{p^n}$ contains $p^n$ elements where $p$ is prime and $n$ is a natural number. For any new theory
there should exist a correspondence principle that at certain conditions the predictions of this theory is
close to the predictions of standard well tested theory. For example, classical theory is a special case
of theory of relativity in the formal limit $c\to\infty$ and a special case of quantum theory in the
formal limit $\hbar\to 0$. Poincare invariant theory is a special case of dS and AdS theories in the formal
limit $R\to\infty$. Analogously, as shown in Refs. \cite{GFQT,JMP,11}, standard theory is a special case
of GFQT in the formal limit $p\to\infty$. In this approach, $p$ is a fundamental quantity defining laws
of physics in our Universe.

One might wonder why we need a new fundamental constant. The history of physics tells us that new 
theories arise when a parameter, which in the old theory was treated as infinitely small or
infinitely large, becomes finite. For example, from the point of view of nonrelativistic physics, the 
velocity of light $c$ is infinitely large but in relativistic physics it is finite. Analogously, from the point
of view of classical theory, the Planck constant $\hbar$ is infinitely small
but in quantum theory it is finite. Therefore it is natural to think that in the future quantum physics 
the quantity $p$ will be not infinitely large but finite.

Since we treat GFQT as a more general theory than standard one, it is desirable not to postulate that GFQT is based on
$F_{p^2}$ because standard theory is based on complex numbers but vice versa, to explain the fact that
standard theory is based on complex numbers since GFQT is based on $F_{p^2}$. Hence, one should find a motivation 
for the choice of $F_{p^2}$ in GFQT. Possible motivations are discussed in Refs. \cite{12,11}, and one of
them is mentioned at the end of this section.

By definition, dS or AdS symmetry in GFQT implies that the operators describing the system under consideration
satisfy the commutation relations (\ref{CR}) which now should be understood as relations in spaces over a
Galois field. Since in GFQT all physical quantities can be only discrete and there are no continuous quantities,
in GFQT all physical quantities are dimensionless and there are no systems of units. This is one of the reasons
why dS and AdS symmetries have a natural generalization to GFQT while Poincare symmetry does not \cite{11}.

In GFQT the notion of probability can be only approximate when $p$ is very large. In particular, the notions
of positive definite scalar product and Hermiticity can be only approximate. In standard theory the difference
between the dS and AdS cases is as follows: Hermitian operators $M^{4\mu}$ in commutation relations (\ref{CR}) for
$\eta^{44}=-1$ become anti-Hermitian when the relations are implemented for $\eta^{44}=1$ and vice versa. 
However, since in
GFQT the notion of Hermiticity can be only approximate, the relations (\ref{CR}) in GFQT can be treated as
the GFQT generalization of dS and AdS symmetries simultaneously. In different situations, a description of
a physical system can be close to a description in standard theory for the dS or AdS cases.

We first discuss an application of GFQT to gravity. It is seen from Eq. (\ref{preNewton2}) that the dS 
correction to standard Hamiltonian disappears if the width of the dS momentum distribution for each body 
becomes very large. In standard theory there is no strong limitation on the width of distribution; 
the only limitation in semiclassical approximation  is that the width of the dS momentum distribution 
should be much less than the mean value of this momentum. Therefore in standard theory the quantities
$\delta_i$ can be very large and then the dS correction practically disappears. As shown in Ref. \cite{gravity},
in GFQT for the validity of the probabilistic interpretation of a wave function, the width of the dS
momentum distribution should be not only much less than $p$ but even much less than $lnp$. Since $p$ is
expected to be a huge number, this should not be a serious restriction for elementary particles. However,
when a macroscopic body consists of many smaller components and each of them is semiclassical, a restriction
on the width of the momentum distribution is stronger when the number of components is greater. This
qualitatively explains that the width of the momentum distribution in the wave function describing a
motion of a macroscopic body as a whole is inversely proportional to the mass of the body. As a consequence,
as noted in Sec. 3, Eq. (\ref{preNewton2}) becomes the Newton law of gravity. A very rough estimation of
the quantity $G$ gives
\begin{equation}
G\approx \frac{R}{m_Nlnp}
\end{equation}
where $m_N$ is the nucleon mass. If $R$ is of order $10^{26}m$ then $lnp$ is of order $10^{80}$ and 
therefore $p$ is of order $exp(10^{80})$. In the formal limit $p\to\infty$ gravity disappears, i.e. in 
our approach gravity is a consequence of finiteness of nature.

Consider now applications of GFQT to particle theory. An elementary particle in GFQT is described by an
IR of the algebra (\ref{CR}) over a Galois field. Consider, for example, how IRs can be constructed in
standard AdS theory. We start from the rest state of a particle (where energy=mass) and gradually construct states with 
higher and higher energies. In such a way, in standard case we obtain the energy spectrum in the range
$[m,\infty)$. However, in the analogous construction in GFQT, we are moving not along a straight line
but along a circumference in Fig. 1 of Ref. \cite{11}. Then sooner or later we will arrive at the point 
where energy=-mass, i.e. at the starting point for constructing an IR for the corresponding antiparticle.
As a consequence, in GFQT one IR describes a particle and antiparticle simultaneously. 
By analogy with the consideration in the preceding section, we now immediately conclude
that in GFQT there are no neutral particles (since it is not possible to reduce the number of states in an IR),
the very notions of a particle and its antiparticle are approximate and such quantum numbers as the electric charge
and the baryon and lepton quantum numbers can be only approximately conserved. All these conclusions are
valid regardless of whether we consider a GFQT analog of dS or AdS theory. 

The fact that in GFQT there are no neutral elementary particles and, in particular, the photon cannot 
be elementary, has been indicated in Ref. \cite{tmf}. As shown in Ref. \cite{FF} titled
"One massless particle equals two Dirac singletons" in standard AdS theory a massless particle can
be composed of two IRs discovered by Dirac in Ref. \cite{DiracS}. As argued in Ref. \cite{11}, in GFQT such
a possibility is even more attractive. 

By analogy with standard theory, the next step is quantization. In the preceding section we discussed two
possibilities. The first one is in the spirit of Dirac's hole theory when a new vacuum is defined by Eq. 
(\ref{newvac}). The problem with this case is that in standard theory, negative energy states contribute
to the energy of the vacuum according to Eq. (\ref{E1}) and the energy becomes a negative infinite number. 
On the other hand, in the approach with the Bogolubov transformation, the new vacuum has zero energy
but symmetry is broken. In GFQT it is broken at huge energies of order $p$ \cite{11} and one might think 
that this is not very important. However, as already noted, it is very desirable not to break symmetry on
quantum level. In GFQT there can be no infinities and, if $p$ is treated only as a cutoff parameter,
one might think that the vacuum energy calculated by analogy with Eq. (\ref{E1}) is of order $p$.
In Galois fields, the notion of positive and negative numbers can be only approximate and a problem
arises what the GFQT analog of Eq. (\ref{E1}) is. This problem has been discussed in Ref. \cite{11}.
The result of calculations is that an analog of Eq. (\ref{E1}) is
\begin{equation}
E_{vac}=\frac{1}{96}(m-3)(s-1)(s+1)^2(s+3)
\end{equation}
where $m$ is the de Sitter mass and $s$ is the spin in units where $\hbar=1/2$. In this units, 
$s=1$ for particles having spin 1/2 in standard theory. Hence for such particles the vacuum energy calculated by
analogy with Eq. (\ref{E1}) is zero. This result demonstrates that $p$ is not only a cutoff parameter
and we have $E_{vac}=0$ instead of $E_1=-\infty$ since the rules of arithmetic in Galois fields are not the same
as in standard mathematics. The result also might  be treated as an indication that only particles with the
spin 1/2 can be elementary. 

In summary, in GFQT it is possible to quantize an IR with the spin 1/2 such that symmetry on quantum level is not
broken and the vacuum energy is zero. This can be achieved in the GFQT analog of Dirac's hole theory.

As noted in the preceding section, the idea of the Bogolubov transformation is that 
creation of a state with the energy $E$ can be described as annihilation of a state with
the energy $-E$. This makes it possible to formally consider a transformation when not only a half but all the
$(a,a^*)$ operators are replaced by the $(b,b^*)$ operators. We call this transformation the AB one. 
A natural requirement is that the operators $M^{ab}$ should be invariant under the AB transformation \cite{11}.
In the usual case the Bogolubov transformation is meaningful only for fermions (see the preceding section).
In GFQT one can express $\eta(i)$ in terms of a constant $\alpha$ such that instead of Eq. (\ref{eta})
\begin{equation}
\alpha {\bar \alpha}=\mp 1
\label{alpha}
\end{equation}
for the normal and broken spin-statistics connection, respectively. As noted in Ref. \cite{11}, the
second possibility is unphysical (not only because the normal spin-statistics connection is broken).
In standard theory the first possibility is impossible but in GFQT, if $p=3\,\,(mod\, 4)$ it is possible 
only if $F_p$ is extended and
the minimum extension is $F_{p^2}$. This can be treated as an argument why standard theory
is based on complex numbers \cite{11}. Also, Eq. (\ref{alpha}) shows that in GFQT both types of statistics
are possible and supersymmetry is not excluded \cite{11}. However, as noted above, if the spin is not 
equal to 1/2 then a problem with the vacuum energy arises.

The above discussion shows that de Sitter symmetry on quantum level gives a new perspective on fundamental notions
of quantum theory.
\begin{center} {\bf Acknowledgements} \end{center}

 The author is grateful to Volodya Netchitailo and Teodor Shtilkind for stimulating discussions.

\end{document}